\documentclass{amsart}

\usepackage{t1enc,amsthm,amssymb,array,epsfig}
\newtheorem{lema}{Lemma}
\newtheorem{prop}[lema]{Proposition}
\newtheorem{teo}[lema]{Theorem}
\newtheorem{rem}[lema]{Remark}

\theoremstyle{definition}
\newtheorem{defin}{Definition}

\newcommand{\R}{\ensuremath{\mathbb{R}}}

\DeclareMathOperator{\Vol}{Vol}
\newcommand{\vol}[1]{\Vol(#1)}

\newcommand{\schr}{Schr\"odinger}
\newcommand{\dis}{\displaystyle}
\newcommand{\Gd}{\mathcal{G}_d}

\newcommand{\Nclass}{\ensuremath{\mathcal{N}_\gamma(\Gd;\R^n)}}
\DeclareMathOperator{\capac}{cap}

\newcommand{\Ho}{{}^{\star} H}
\newcommand{\cH}{\mathcal{ H}}

\newcommand{\p}{\partial}
\newcommand{\imp}{\Rightarrow}

\begin{document}
\date{compiled \today}

\title[Discrete spectra of $p$-branes and $M5$ brane]{The discrete spectrum of the $D=11$ bosonic $M5$-brane.}
\author[I. Martin, L. Navarro, A. J. P\'erez A. and A. Restuccia]{I. 
Martin $^1$, L. Navarro $^2$, A. J. P\'erez A.$^3$ and A. Restuccia$^4$}
\address{$^{1}$ Departamento de F\'\i sica, Universidad Sim\'on Bol\'\i
var, Apartado 89000, Caracas 1080-A, Venezuela. }
\email{isbeliam@usb.ve}
\address{$^{2}\,$ Departamento de Matem\'aticas, Universidad Sim\'on Bol\'\i
var, Apartado 89000, Caracas 1080-A, Venezuela.}
\email{lnavarro@ma.usb.ve}
\address{$^{3}\,$ Departamento de Matem\'aticas, Universidad Sim\'on Bol\'\i
var, Apartado 89000, Caracas 1080-A, Venezuela.}
\email{ajperez@ma.usb.ve}
\address{$^{4}$ Departamento de F\'\i sica, Universidad Sim\'on Bol\'\i
var, Apartado 89000, Caracas 1080-A, Venezuela. }
\email{arestu@usb.ve} \maketitle

\begin{abstract} %\ %\newline
    We prove that the spectrum of the regularized $M5$-brane in 
    $D=11$ target space is discrete with eigenvalues
    extending to $\infty$. The proof includes the same result for the 
    spectra of regularized bosonic $p$-branes in general.

\end{abstract}

%\maketitle

\vskip 1 cm
\section{Introduction}
The understanding of the spectral properties of the supermembrane and 
super $M5$-brane in 11 dimensions are important steps towards the 
non-perturbative analysis of M-theory. The $SU(N)$ regularized 
Hamiltonian of the supermembrane on a $D=11$ Minkowski target space has 
a continuous spectrum \cite{dwln}, see also \cite{helling}, 
\cite{lucher}, \cite{dwnh}, \cite{dwmn}. The  supermembrane on a $D=11$ 
target space with a compact sector is expected to have also a continuous spectrum 
\cite{dwpp}. But , the $D=11$ supermembrane wrapped in an 
irreducible way on a compact sector of the target space, i.e., with 
a topological condition on configuration space yielding a non trivial 
central charge has a discrete spectrum and its ground state has a 
strictly positive energy \cite{gmr}, \cite{lbar}, \cite {bgmmr}, \cite {bgmr}, \cite{br}, \cite {Martin:2001zv}. 

In all cases the bosonic Hamiltonian has a discrete spectrum. However, they are 
qualitatively different in a crucial way. In the latter case, the 
central charge generate mass terms implying that the potential on 
configuration space tends to infinity when it approaches infinity in 
this space. Moreover, this qualitative property of the spectrum 
remains unchanged in the supersymmetric theory with non trivial 
central charges. 
In the former case \cite{dwln}, the potential 
presents zero point valleys extending to infinity on configuration 
space. In this case the membrane admits as physical configurations 
stringy spikes that make the supermembrane spectrum continuous, 
despite the fact that its bosonic  part on its own would not produce a 
continuous spectrum. Although the bosonic potential is zero on the minima of the valleys, 
the walls of the valleys get closer as they approach infinity in a 
way that the quantum mechanical wave function cannot escape to 
infinity. The precise mathematical meaning of this property was 
explained in \cite {gmnpr}. It is formulated in terms of the 
integral of the potential on a fixed sized cell, defined in the sense 
of Molchanov and Maz'ya and Shubin \cite{molchanov}, \cite{Maz-Shu}, when the center of the cell 
approaches infinity on configuration space. The potential integral of 
the cell in the directions of zero potential is bounded below by the 
potential of an harmonic oscillator ensuring that the integral goes 
to infinity when the cell is moved to infinity in those directions.
This bound from below is lost once the fermionic part of the 
potential is added, as a consequence the supermembrane spectrum 
becomes continuous. All the above results refer to regularized 
Hamiltonians.
More recently, it has been shown  \cite {bgmr}, \cite {bgmr1} that the exact bosonic 
Hamiltonian for the case of the supermembrane with non trivial 
charges, has a discrete spectrum. This was achieved with a precise 
definition of the configuration space in terms of Sobolev spaces. It 
was also proven that the spectrum of the  $SU(N)$ regularized model  of 
the semiclassical Hamiltonian converges to the spectrum of the exact 
Hamiltonian when N tends to infinity.

In the case of the $M5$-brane no results have been reported concerning 
its spectrum. In \cite{msjvr} a semiclassical analysis of the 
spectrum of the $M5$-brane was performed. The $M5$-brane covariant action was first obtained in 
\cite {Pasti1}, \cite {Pasti2} and a  gauge fixed action version was obtained independently 
in \cite {JS3}. In \cite{DeCastro:2001gp} a formulation was obtained for its  Hamiltonian in 
terms of first class constraints only. Also, its BRST structure and the 
existence of string and membrane spikes were shown. Later, the 
Nambu-Poisson structure of the $M5$-brane was introduced in 
\cite {DeCastro:2004}. Also,  a general analysis of such structure for 
$p$-branes was analyzed in \cite {hoppe}.
In this paper, following \cite {DeCastro:2004}, \cite {gmnpr} we 
prove that the spectrum of the regularized $M5$-brane and that one of any $p$-brane 
is discrete. The proof in general applies to 
many matrix models associated to such theories. In section 2, the 
algebraic structure of the $M5$-brane Hamiltonian is presented for 
completeness.
 In section 3, we present the $SU(N)$ regularized version of the 
 $M5$-brane Hamiltonian exploiting the Nambu-Poisson structure 
 underlying it. In section 4,  using appropriate theorems of 
 spectral analysis, we show that generally  $p$-branes matrix models present discrete spectra. 
 With this result at hand, the discreteness of the spectrum of the regularized bosonic $M5$-brane 
 Hamiltonian follows easily. In section 5, we present conclusions.
%-----------------------------------------------------------------------------
\section{The Algebraic Structure of M5-Brane Hamiltonian}\label{Hamiltonian}

We start recalling the bosonic M5-Brane Hamiltonian on a $D=11$ 
Minkowski target space in the
light cone gauge  obtained in \cite{DeCastro:2001gp},
\begin{equation}\label{hp}
{\cH}=\frac{1}{2}\Pi^M\Pi_M+2g+l^{\mu \nu}l_{\mu\nu}+\Theta_{5i}\Omega^{5i}+\Theta_j\Omega^j
+\Lambda^{\alpha\beta}\Omega_{\alpha\beta}
\end{equation}
where

\begin{equation}\label{l}
l^{\mu\nu}=\frac{1}{2}(P^{\mu\nu}+{\Ho}^{\mu\nu})
\end{equation}
and
\begin{eqnarray}
{\Ho}^{\mu\nu}&=& \frac{1}{6} \epsilon^{\mu\nu\gamma\delta\lambda}
H_{\gamma\delta\lambda}\\
H_{\gamma\delta\lambda}&=& \p_{\rho}B_{\lambda\sigma}
+\p_{\sigma}B_{\rho\lambda} + \p_{\lambda}B_{\sigma\rho}
\end{eqnarray}

$\Theta_{5i}$, $\Theta_j$, $\Lambda^{\alpha\beta}$ are the
Lagrange multipliers associated to  remaining constraints
\begin{eqnarray}
\Omega^{5i}&=&P^{5i}-{\Ho}^{5i}=0\label{fclass1}\\
\Omega^{j}&=&\partial_{\mu}P^{\mu j}=0\label{fclass2},\quad
i=1,2,3,4
\end{eqnarray}
\begin{equation}\label{fclass3}
\Omega_{[\alpha\beta]}=\partial_{[\beta}\left[
\frac{1}{\sqrt{W}}(\Pi_M\partial_{\alpha]}{X}^M+\frac{1}{4}V_{\alpha]})\right]=0
\end{equation}
where
\begin{equation}
V_\mu=\epsilon_{\mu\alpha\beta\gamma\delta}
l^{\alpha\beta}l^{\gamma\delta}
\end{equation}

$\Pi^M$ and $P^{\mu\nu}$ are the canonical conjugate momenta to ${X}^M$ 
and $B_{\mu \nu}$ respectively, $g$ is the determinant of the induced 
metric.
Equations (\ref{fclass1}) and (\ref{fclass2}) are the first class
constraints  generating the local gauge symmetry associated to the
antisymmetric field while  (\ref{fclass3}) is  the first class
constraint generating volume preserving diffeomorphisms. ${X}^M$ are the light cone 
gauge transverse coordinates on the target space.  

$W$ is a scalar density introduced in the gauge fixing procedure. It represents the determinant
of an intrinsic metric over the spatial world volume of the brane. In our
notation caps Latin letters are transverse light cone gauge
indices $M,N=1,\ldots,9$,  Greek ones are spatial world volume
indices, and small Latin letters denote spatial world volume
indices on a $4$-dimensional spatial submanifold.

The elimination of second class constraints from the formulation in 
\cite{Pasti1}, \cite{Pasti2} and \cite{JS3} to produce a canonical Hamiltonian with
only first class constraints, was achieved at the price of loosing the
manifest 5 dimensional spatial covariance. In this way, the spatial
world volume splits into $M_{5}= M_{4}\times M_{1}$. We will exploit
that decomposition in our analysis of the Hamiltonian. We will assume 
$M_{4}$ has a symplectic structure with $\omega ^{0}$ being its 
associated non degenerate closed $2$-form. It is assumed that $M_{4}$ and 
$M_{1}$ are compact manifolds.

Let us analyze the Hamiltonian density  term by term. We first notice that $g$, the
determinant of the induced metric, may be re-expressed in a
straightforward manner as a squared five entries bracket, a 
Nambu-Poisson bracket,
\begin{eqnarray}
g&=&\frac{1}{5!}\epsilon^{\nu_1,\ldots,\nu_5}\epsilon^{\mu_1,\ldots,\mu_5}g_{\mu_1\nu_1}\ldots
g_{\mu_5\nu_5}\cr\cr &=&\frac{1}{5!}\{X^M,X^N,X^P,X^Q,X^R\}^2.
\end{eqnarray}

Let us consider now the  third term dependent on the antisymmetric 
field $B_{\mu\nu}$. It
is invariant under the action of the first class constraints (\ref{fclass1})
and (\ref{fclass2}). To eliminate part of these constraints, we proceed
to make a partial gauge fixing on $B_{\mu\nu}$, following \cite{DeCastro:2001gp} we take
\begin{equation}
B_{5i}=0
\end{equation}
which, together with the constraint (\ref{fclass2}) allow us a
canonical reduction of the Hamiltonian (\ref{hp}). Notice that the
contribution of this partial gauge fixing to the functional
measure is $1$.  We are then left with the constraint
\begin{equation}\label{fclass4}
\p_j P^{ij}+\p_5{\Ho}^{5i}=0 \quad\quad i,j=1,2,3,4
\end{equation}
which generates the gauge symmetry on the 2-form $B$

\begin{equation}
\delta B_{ij}=\p_i\Lambda_j-\p_j\Lambda_i
\end{equation}
$B$ as a $2$-form over $M_4$ may be  decomposed using the Hodge decomposition
theorem into an exact form plus a co-exact form plus an harmonic
form. Its exact part is canonically conjugate to the co-exact part of 
$P^{ij}$, that is, calling $\p_i \dot{b_{j}}$ the exact part of $B_{ij}$ we 
have
\begin{equation}
 \langle {P^{ij}\p_i \dot{b_{j}}}\rangle =  \langle { -\p_iP^{ij} \dot{b_{j}}}\rangle
\end{equation}
then an admissible gauge fixing is to set $b_{j}=0$ to
eliminate the exact form and the  $\p_iP^{ij}$ from the constraints.

We are then left with the co-exact part of $B$. It is directly related 
to $l^{5i}$ 
\begin{equation}
l^{5i}=\epsilon^{ijkl}\partial_{j}B_{kl}.
\end{equation}

Noticing that $l^{5i}$ is divergenceless, it may always be rewritten 
without loosing generality as
\begin{equation}\label{decom2}
l^{5i}=\epsilon^{5ijkl}\partial_j\phi_{[a}\partial_k\phi_b\partial_l\phi_{c]},
\quad\quad a,b,c=1,2,3.
\end{equation}

This decomposition in terms of scalars is always valid locally for any
four dimensional divergenceless smooth vectorial density. $\phi_{a}$, 
$a=1,2,3$ represent the three degrees of freedom of the 
co-exact part of $B_{ij}$.

Now we decompose the tensor density $l^{ij}$ into
\begin{equation}\label{decom1}
l^{ij}=\epsilon^{ji\alpha\beta\gamma}
\partial_\alpha\phi_{[a}\partial_\beta\phi_b\partial_\gamma\phi_{c]}+\epsilon^{jikl}\omega_{kl}
\end{equation}
where  $\omega$ is a closed $2$-form. 

It is now possible, following the Darboux's theorem, to express
$\omega_{kl}$ in terms of the canonical $2$-form $\omega^{0}$ over
$M_{4}$.
In fact the area preserving diffeomorphisms homotopic to the identity 
are generated by $\Omega_{\alpha \beta }$ with infinitesimal 
parameter $\xi^{\alpha \beta }$ or equivalently generated by
\begin{equation}
\Omega_{\alpha }=\left(\Pi_M\p_{\alpha}{X}^M+\frac{1}{4}V_{\alpha}\right)
\end{equation}
with infinitesimal parameter $\xi^{\alpha}$ given by  
\begin{equation}
\xi^{\alpha  }= \frac{1}{\sqrt{W}}\p_{\beta}( \sqrt{W}\xi^{\alpha \beta })
\end{equation}
satisfying identically
\begin{equation}
\p_{\alpha}( \sqrt{W}\xi^{\alpha })=0.
\end{equation}

This volume preserving restriction leaves the four spatial parameters 
$\xi^{\alpha }$ associated to $M_{4}$ unconstrained.
So, we are allowed to use the Darboux procedure to fix $\omega$ to 
$\omega ^{0 }$. We should be left still with one free parameter since 
the local degrees of freedom of $\omega$ are only three. Indeed, that 
is the case since the corresponding gauge fixing procedure allows to 
eliminate $l^{5i}$ from the constraints in the following way,

From (\ref {fclass3}) we have 
\begin{equation}
 \frac{1}{\sqrt{W}}\left(\Pi_M\p_{\alpha}{X}^M+\frac{1}{4}V_{\alpha}\right)=\p_{\alpha} U
\end{equation}
where $U$ is an auxiliary field. We notice in $V_{\alpha}$, the 
product of the $\phi$ dependent terms is zero. In particular,
\begin{equation}
    V_{i}= - 4l^{5j} \omega^{0 }_{ij}
\end{equation}
it allows to eliminate $l^{5j }$ from the equation for $\alpha =i$ in 
terms of $U$, which is determined from the equation for $\alpha =5$:
\begin{equation}
 l^{5j}= -\epsilon ^{jikl}\omega^{0 }_{kl}\p_{i}U +  \frac{\epsilon 
 ^{jikl}\omega^{0 }_{kl}\Pi_{M}\p_{i}X^{M}}{\sqrt{W}}
\end{equation}
we are then left with the remaining constraint 
\begin{equation}
 \p_{j}l^{5j }= \epsilon ^{jikl}\omega^{0}_{kl}\p_{j}\left(\frac{\Pi_{M}\p_{i}X^{M}}{\sqrt{W}}\right)=0.
 \end{equation}

 The kinetic term associated to this gauge fixing  is a total 
 time derivative and, since $\omega^{0}$ is time independent, it can 
 be eliminated from the action. 
 Finally, we are left with a complicated $l^{\mu\nu}l_{\mu\nu}$ term 
 but to prove the discreteness of the spectrum it will become 
 irrelevant as we will see in the following sections. The final gauge 
 fixing corresponding to the symplectomorphisms preserving 
 $\omega^{0}$ is performed by taking the Lagrange multiplier of the 
 associated first class constraint to be zero, the ghosts fields 
 decouple from the action.

\section{Regularization of the M5-brane}\label{matricial}

After fixing $\omega$ to $\omega^{0}$ we may resolve the
volume-preserving constraint for $\phi^a$ $a=1,2,3,4$. We are then
left still with one constraint,
\begin{equation}\label{fcc}
\epsilon^{ijkl}\omega^0_{kl}\p_i\left(\frac{\Pi_M\p_jX^M}{\sqrt{W}}\right)=0.
\end{equation}

The left hand member generates the symplectomorphisms preserving
$\omega^0$. The full five dimensional diffeomorphisms have been
reduced to only that generator. We are then left with a
formulation in terms of $X^M$ and its conjugate momenta $\Pi_M$,
invariant under symplectomorphisms. The antisymmetric field
$B_{\mu\nu}$ and its conjugate momenta $P^{\mu\nu}$ have been
reduced to $\omega^0$, there is no local dynamics related to them. All
the dynamics may be expressed in terms of $(X^M,\Pi_M)$. 

In order to obtain a regularization of the  Hamiltonian, we 
express $X^M(\tau,\sigma)$ and $\Pi_M$ in terms of a complete 
orthonormal basis over $M_{4} \times M_{1}$, 
$\{Y_{a}(\sigma^{1},\sigma^{2},\sigma^{3},\sigma^{4})\}$, in the Hilbert 
space of $L^{2}$ functions  for $M_{4}$ and a Fourier basis for the 
$M_{1}$ manifold.  
\begin{eqnarray}
 X^M(\tau,\sigma)=X^{a\:M}(\tau)Y_{a}(\sigma)\quad  a= 1,2,\ldots,\infty \\
 \Pi_{M}(\tau,\sigma)= \sqrt{W}\Pi^{a}_{M}(\tau)Y_{a}(\sigma).
\end{eqnarray}

Since for every pair $a,b$ 
\begin{equation}
\frac{\epsilon^{ijkl}\omega^0_{kl}}{\sqrt{W}}\p_iY_a\p_jY_b
\end{equation}
is a scalar function over $M_4$, we may reexpress it in terms of
the basis, and obtain the symplectic bracket
\begin{equation}
\{Y_a,Y_b\}=\frac{\epsilon^{ijkl}\omega^0_{kl}}{\sqrt{W}}\p_iY_a\p_jY_b=f_{abc}Y_c
\end{equation}
where $f_{abc}$ is  given by
\begin{equation}
\int_{M_4}\frac{\epsilon^{ijkl}\omega^0_{kl}}{\sqrt{W}}\p_iY_a\p_jY_b Y_c=f_{abc}
\end{equation}
and satisfy the Jacobi identity. These are the structure constants of 
the symplectomorphisms preserving $\omega^{0}$. 
Furthermore, we may introduce
\begin{equation}
Y_aY_b=C_{abd}Y_d
\end{equation}
it is again valid since $Y_a Y_b$ is also a scalar function
over $M_4$. We get
\begin{equation}
\int_{M_4}Y_aY_bY_d=C_{abd}
\end{equation}
which becomes totally symmetric in $a,b,d$.
The other natural bracket in the formalism \cite {DeCastro:2004}
is the Nambu one,
\begin{equation}    
\{A,B,C,D,E\}=\frac{1}{\sqrt{W}}\epsilon^{\alpha\beta\gamma\delta\rho}\p_{\alpha}A\p_{\beta}B\p_{\gamma}C\p_{\delta}D\p_{\rho}E
\end{equation}
in particular the scalar
\begin{equation}
 \{Y_a,Y_b,Y_c,Y_d,Y_e\}=f^{g}_{abcde}Y_{g}
\end{equation}
where
\begin{equation}
 \int_{M_5}\{Y_a,Y_b,Y_c,Y_d,Y_e\}Y_{g}=f_{abcdeg}
\end{equation}
is a totally antisymmetric tensor satisfying a generalized Jacobi 
identity \cite{Nambu:1973qe}, \cite{Takhtajan:1994vr}, \cite{Grabowski1}, \cite{Grabowski2}. By construction we have the following relation for 
the compact base manifold we are considering
\begin{equation}
 f^{g}_{abcde} = \sum_{\textstyle{antisymm} (a,b,c,d,e)} 
 in_{a}f^{\hat{c}}_{bc}f^{\hat{e}}_{de}C^{h}_{\hat{c}\hat{e}}C^{g}_{ha}
    \label{struct5}
\end{equation}
where $-n^{2}_{a}$ is an eigenvalue of the Laplacian on $M_{1}$. The 
right hand side satisfies the generalized Jacobi identity by 
construction.

We now consider a regularization of the $M5$-brane Hamiltonian by 
truncation of the infinite dimensional basis, that is, $a=1,2\ldots.N$. We require in addition 
that in the remaining configuration space there exists brackets to 
have an intrinsic definition of the parameters $f^{c}_{ab}$ and 
$C_{abc}$ entering in the theory. In the large $N$ limit the corresponding structure 
constants should be the area preserving ones. If so, we will have in 
the large $N$ limit a generalized Jacobi identity for 
$f^{g}_{abcde}$. Meanwhile, it is not necessary to require a 
generalized Jacobi identity for the truncated theory. 
In the discreteness proof presented here we do not use any algebraic 
properties of the brackets. It is valid for any truncation in terms 
of some constants $f^{g}_{abcde}$. However, if we require an intrinsic 
meaning for the truncated theory, the algebraic structure should be 
present.
We notice that the Nambu structure constants $f^{g}_{abcde}$ of the 
symplectomorphisms satisfy the following properties
\begin{equation}
\int_{M_{5}} |\{Y_a,Y_b,Y_c,Y_d,Y_e\}|^{2} = f_{abcdeg}f^{abcdeg} > 0
\nonumber
\end{equation}
and more generally
\begin{equation}
M^{gh}=f^{g}_{abcde}f^{abcdeh}
\end{equation}
is a strictly positive matrix.
This is the only assumption we will require on the truncated theory 
for the associated $f^{g}_{abcde}$.

All the interacting terms in the Hamiltonian density  may now be rewritten using 
${X}^{a\:M}(\tau)$, their conjugate momenta and the 
structure constants $f_{abc}$ and $C_{abd}$. Integrating on the 
spatial coordinates we arrive to a quantum mechanical Hamiltonian. 

For the purpose of analyzing the spectrum of this regularized 
Hamiltonian, we will concentrate on the first interacting terms that 
do not include $l^{\mu\nu}$, this procedure is perfectly justified since  

\begin{equation}\label{hp2}
{\cH}=\frac{1}{2}\Pi^M\Pi_M+2g+l^{\mu \nu}l_{\mu\nu}\geq 
\frac{1}{2}\Pi^M\Pi_M+2g = {\cH}_{0}
\end{equation}
so, in what follows the spectrum of ${\cH}_{0}$ is studied. We will prove that ${\cH}_{0}$ has a discrete spectrum $\lambda_{n}$, with 
$\lambda_{n} \rightarrow \infty$ when $n \rightarrow \infty $. The 
min max theorem assures the same qualitative spectrum for ${\cH}$ as 
it will be shown in the next section. The regularized Hamiltonian 
density ${\cH}_{0}$ will correspond to a $p=5$-brane.

\section{The spectrum of the quantum mechanical Hamiltonian for 
regularized $p$-brane potentials}

Let us consider the \schr\ operator 
\begin{equation}
    H_{L}= -\Delta+V_{L}(X)=-\Delta+(X^{a_{1}}_{M_{1}}\ldots 
    X^{a_{L}}_{M_{L}}f_{a_{1}\ldots a_{L}}^{b})^{2}
\end{equation}
where $L$ is the degree of the brane considered, $M_{i}= 1,\ldots,K$, 
$a_{i}=1,\ldots,N$, $K\ge L$, $N\ge L$, $X=X^{a}_{M} \in \R^{KN}$ and
$f_{a_{1}\ldots a_{L}}^{b}$ is a constant tensor totally 
antisymmetric in ${a_{1},\ldots, a_{L}}$ and it is not singular, i.e., 

\begin{equation}
 \textrm{if}\quad X^{a_{1}}_{M}f_{a_{1},a_{2}\ldots a_{L}}^{b} =0 \quad\textrm{then}\quad  
 X^{a_{1}}_{M}=0.
\end{equation}

We introduce 
\begin{equation}
    H_{l}= -\Delta+V_{l}(X)\qquad 1\leq l \leq L
   \end{equation}
where
\begin{eqnarray}
V_{l}=\sum_{{\small\begin{array}{cc}
M_{1} \neq M_{2} \neq \ldots \neq M_{l}\\
1 \leq  M_{i}  \leq  K
\end{array}}}
(X^{a_{1}}_{M_{1}}\ldots 
    X^{a_{l}}_{M_{l}}f_{a_{1}\ldots a_{l}b_{l+1}\ldots b_{L}}^{b}) (X^{\hat{a_{1}}}_{M_{1}}\ldots 
    X^{\hat{a_{l}}}_{M_{l}}f_{\hat{a_{1}}\ldots \hat{a_{l}} 
    b_{l+1}\ldots b_{L}}^{b}).  \nonumber 
    \end{eqnarray}

In this section we prove that $H_{l}$, $ 1 \leq  l \leq L$   has a discrete spectrum, but first let us recall some mathematical definitions and theorems needed.

The definition of capacity of a compact set in $\R^n$ is an important 
ingredient in the Maz'ya and Shubin generalization of Molchanov's 
theorem \cite{molchanov} on the necessary and sufficient conditions for the discreteness of the 
spectrum of the \schr\ operator. For details and the full-general 
version of the Maz'ya and Shubin theorem see \cite{Maz-Shu}.
\begin{defin}
Let $n\ge3$, $F\subset\R^n$ be a compact, and $Lip_{c}(\R^n)$ the set of all real-valued functions with compact support satisfying a uniform Lipschitz condition in $\R^n$. The Wiener's capacity of $F$ is defined by
\[\capac(F)=\capac_{\R^n}(F)=\inf\left\{\int_{\R^n}|\nabla u(x)|^2\,dx\Big|\ u\in Lip_{c}(\R^n),\ u|_F=1\right\}.\]
\end{defin}

In physical terms the capacity of the set $F\subset\R^n$ is defined as the
electrostatic energy over $\mathbb{R}^{n}$ when the electrostatic
potential is set to $1$ on $F$.\\

\begin{defin}
Let $\Gd\subset\R^n$ be an open and bounded star-shaped set of diameter $d$, let $\gamma\in(0,1)$. The \emph{negligibility class} \Nclass\
consists of all compact sets $F\subset\overline{\Gd}$ satisfying
$\capac(F)\leq\gamma\capac(\overline{\Gd})$.
\end{defin}

For example we can take $\Gd$ to be a n-cube or a ball in $\R^n$. In what follow we refer to $\Gd\setminus F$ as a cell.

\begin{teo}[Maz'ya and Shubin]\label{teo_mazya}
Let $V\in L^1_{\text{loc}}(\R^n)$, $V\geq0$.

Necessity: If the spectrum of $-\Delta+V$ in $L^2(\R^n)$ is discrete
then for every function $\gamma:(0,+\infty)\rightarrow(0,1)$ and
every $d>0$
\begin{equation}\label{inf_int}
\inf_{F\in\Nclass}\int_{\Gd\setminus
F}V(x)\,dx\rightarrow+\infty\quad\text{as}\quad
\Gd\rightarrow\infty.
\end{equation}

Sufficiency: Let a function $d\mapsto\gamma(d)\in(0,1)$ be defined for $d>0$ in a neighborhood of $0$ and satisfying
\[ \limsup_{d\downarrow0}d^{-2}\gamma(d)=+\infty. \]

Assume that there exists $d_0>0$ such that
\emph{(\ref{inf_int})} holds for every $d\in(0,d_0)$. Then the
spectrum of $-\Delta+V$ in $L^2(\R^n)$ is discrete.
\end{teo}

\begin{rem}
It follows from the previous theorem that a necessary condition
for the discreteness of spectrum of $-\Delta+V$ is
\begin{equation}\label{necessary}
\int_{\Gd}V(x)\,dx\rightarrow\infty
\quad\text{as}\quad\Gd\rightarrow\infty.
\end{equation}
\end{rem}

Let us recall that  K. Friedrichs (see \cite{Maz-Shu} for further references) proved that the spectrum of the
\schr\ operator $-\Delta+V$ in $L^2(\R^n)$ with a locally
integrable potential $V$ is discrete provided
$V(x)\rightarrow\infty$ as $|x|\to\infty$.

In what follow we will denote the $n$-dimensional Lebesgue measure by $\vol{\cdot}$.

\begin{lema}\label{lema}
For each given ball $\Gd=\Gd(x_0)$ centered at $x_0$ and radius $d>0$.
\[\inf_{F\in\Nclass}\vol{\Gd\setminus F}>0. \]
\end{lema}
\begin{proof}
Let $V(x)=|x|$. Then by Friedrichs theorem the spectrum of
$-\Delta+V$ is discrete, so by Theorem \ref{teo_mazya} we have
\[ \inf_{F\in\Nclass}\int_{\Gd\setminus F}V(x)\,dx\to\infty\quad\text{as}\quad |x_0|\to\infty. \]
Now $\int_{\Gd\setminus F}V(x)\,dx\leq(|x_0|+d)\vol{\Gd\setminus
F}$ implies that
\[ \inf_{F\in\Nclass}\int_{\Gd\setminus F}V(x)\,dx \leq (|x_0|+d)\inf_{F\in\Nclass}\vol{\Gd\setminus F} \]
from which follows that $\inf_{F\in\Nclass}\vol{\Gd\setminus
F}>0$, as we claimed.
\end{proof}

The min-max principle is useful in the proof of the next proposition, so we state it 
for completeness.
\begin{teo} [Min-max principle]\label{min}

    Let $H$ be a self-adjoint operator that is bounded from below, i.e, $H\geq cI$ for some $c$. 
Define 
\begin{equation}\nonumber 
\mu_{n}(H)= \sup_{\phi_{1},\phi_{2}\ldots\phi_{n-1}}U_{H}(\phi_{1},\phi_{2}\ldots\phi_{n-1}) 
\end{equation}
where
\begin{equation}\nonumber  
U_{H}(\phi_{1},\phi_{2}\ldots\phi_{m})=\inf (\psi, H \psi) \quad \textstyle{when}\quad  \|\psi\|=1 \quad\textstyle{and} \quad\psi \in \left[\phi_{1},\phi_{2}\ldots\phi_{m}\right]^{\perp}
\end{equation}
	   
$\left[\phi_{1},\phi_{2}\ldots\phi_{m}\right]^{\perp}$ is shorthand 
for $\{\psi| (\psi,\phi_{i})=0, 
	   i=1,2,\ldots,m\}$. The $\phi_{i}$ are not necessarily 
	   independent. 
	   
	   Then, for each fixed $n$, either:
	   
	   (a) there are $n$ eigenvalues (counting degenerate 
	   eigenvalues a number of times equal to there multiplicity) 
	   below the bottom of the essential spectrum, and 
	   $\mu_{n}(H)$ is the $n$th eigenvalue counting multiplicity;
	   
	   or
	   
	   (b) $\mu_{n}$ is the bottom of the essential spectrum, i.e., $\mu_{n}=\inf\{\lambda|\lambda\in \sigma_{ess}(H)\}$, and in that case $\mu_{n}=\mu_{n+1}=\mu_{n+2}=\ldots$ and there 
	   are at most $n-1$ eigenvalues (counting multiplicity) 
	   below $\mu_{n}$.
    \end{teo}
    
    For proof of the min-max theorem and further reading on the 
    subject see \cite{reed}.
    
    \begin{rem}
	  As a consequence of this theorem if $A$ and $B$ are self-adjoint 
    operators bounded from below and if $A\leq B$, then 
    $\mu_{n}(A)\leq \mu_{n}(B)$. In this case if 
    $\mu_{n}(A)\rightarrow\infty$ when $n\rightarrow\infty$ then 
    $\mu_{n}(B)$ will also tend to infinity. This, in turn, means 
    that if $A$  has a discrete spectrum with a compact resolvent 
    then $B$ will also have a discrete spectrum with a compact 
    resolvent. 
   \end{rem}

Now, we prove a theorem concerning the above defined operator $H_{l}$. In what follow we use DS for discrete spectrum.

 \begin{prop}\label{seq}
Let
\[V_{l}=\sum_{1 \leq  M_{i}  \leq  K}(X^{a_{1}}_{M_{1}}\ldots X^{a_{l}}_{M_{l}}f_{a_{1}\ldots a_{l}b_{l+1}\ldots b_{L}}^{b})^2\]
with $M_{1} \neq M_{2} \neq \ldots \neq M_{l}$.

Then the following sequence holds:

	\begin{center} $H_{l}$ has DS $\imp$ $-\Delta+\sqrt{V_{l}}$ has DS $\imp$ $H_{l+1}$ has DS.\end{center}

\end{prop}

\begin{proof}
%%%%%%%%%%%%%%%%%%%%%%%%%%%%%%%%%%%%%%%%%%%%%%%%%%%%%%%%%%%%%%%%%%%%%%%%%%%%%%%%%%%%%%%%%%%%%%%%%%%%
	{\bf A.} \textit{$H_{l}$ has DS $\imp$ $-\Delta+\sqrt{V_{l}}$ has DS}.
\vskip 0.15cm
Let $\Gd=\Gd(X_0)\subset\R^{KN}$ be a ball centered at $X_0$ and radius $d>0$, let $F\in\Nclass$. Then $X=X_{0}+\xi$ for all $X$ in the cell $\Gd\setminus F$. Let $\Omega_F$ be the set of all such $\xi$. Then the necessary 
	condition of Theorem \ref{teo_mazya} implies that
	\begin{equation}\label{cell}	    
	  \inf_{F}\int_{\Omega_F}  
	  V_{l}(X_0,\xi)\,d\xi\, {\rightarrow  \infty} \quad\text{as}\quad{|X_0|\rightarrow 
	  \infty}
	\end{equation}

We can rewrite the potential as 
\begin{equation}\label{eq0}
\dis V_l=\sum_{j=1}^{N_l}P^2_j(X_0, \xi)
\end{equation}

where $P_j,\ j=1,\ldots, N_l$, are polynomials in $\xi$ with 
coefficients depending on $X_0$. Using the Gram-Schmidt process it is 
possible to rewrite 
\[P_j(X_0,\xi)=\sum_{k=1}^{n_j}a_{jk}(X_0)\varphi_{jk}(\xi)\] where $\{\varphi_{jk}(\xi)\}$ is a finite system of orthonormal polynomials depending on $\Omega_F$ i.e., 

\[\int_{\Omega_F}\varphi_{jk}(\xi)\varphi_{im}(\xi)\, d\xi=\delta_{(j,k)(i,m)}\quad (\textrm{the Kroneker delta}).\]
and $a_{jk}(X_0)$ are its corresponding coefficients.
\vskip 0.3cm
Note that for any system $\{\varphi_{jk}\}$ there exists $M_F>0$ such 
that $|\varphi_{jk}(\xi)|\le M_F$ for all $(j,k)$ and all $\xi\in\Omega_F$.

Hence
\begin{equation}\label{eq1}
\int_{\Omega_F}P_j^2(X_0,\xi)=\sum_{k=1}^{n_j}a^2_{jk}(X_0)_{\Omega_F}=:\|P_j\|^2_{\Omega_F}\textrm{ and }\int_{\Omega_F}V_l\,d\xi=\sum_{j}\|P_j\|_{\Omega_F}^2
\end{equation}

It is possible to choose $N_0\ge\max\{N_{l}, n_j:j=1,\ldots, N_l\}$ 
independent of $F$, then using  $\dis \left(\sum_{k=1}^na_k\right)^2\le n\sum_{k=1}^na_k^2$ twice, we have $\dis P_j^4\le N_0^3\sum_{k=1}^{n_j}a_{jk}^4\varphi^4_{jk}(\xi)$, therefore
\[\int_{\Omega_F}P_j^4\,d\xi\le 
N_0^3\sum_{k=1}^{n_j}a^4_{jk}\int_{\Omega_F}\varphi^4_{jk}(\xi)\,d\xi\le N_0^3M_F^2\sum_{k=1}^{n_j}a_{jk}^4\le N_0^3M_F^2\left(\sum_{k=1}^{n_j}a_{jk}^2\right)^2\]
i.e. $\dis \int_{\Omega_F}P_j^4\,d\xi\le N_0^3M_F^2\|P_j\|_{\Omega_F}^4$. Then from this and (\ref{eq0}) and (\ref{eq1})
\begin{equation}\label{eq4}
\int_{\Omega_F}V_l^2(X_0,\xi)\,d\xi\le N_0^4M_F^2\sum_{j}\|P_j\|_{\Omega_F}^4\le N_0^4M_F^2\left(\int_{\Omega_F}V_l(X_0,\xi)\,d\xi\right)^2
\end{equation}
Because of $V_l^{\alpha}\in L^2(\Omega_F)$ for all $\alpha\ge0$, using Schwarz inequality  twice we obtain:
\begin{equation}\label{eq5}
\left(\int_{\Omega_F}V_l\,d\xi\right)^{3/2}\le\int_{\Omega_F}V_l^{1/2}\,d\xi\left(\int_{\Omega_F}V_l^2\,d\xi\right)^{1/2}
\end{equation}
Now using (\ref{eq4}) and (\ref{eq5}) we have:
\[\left(\int_{\Omega_F}V_l\,d\xi\right)^{1/2}\le N_0^2M_F\int_{\Omega_F}V_l^{1/2}\,d\xi\]
And from (\ref{cell}) we concluded that \[\inf_{F\in\Nclass}M_F>0 \textrm{ and }-\Delta+\sqrt{V_l} \textrm{ has DS.}\]

%%%%%%%%%%%%%%%%%%%%%%%%%%%%%%%%%%%%%%%%%%%%%%%%%%%%%%%%%%%%%%%%%%%%%%%%%%%%%%%%%%%%%%%%
\vskip 0.5cm
{\bf B.} \textit{$-\Delta+\sqrt{V_{l}}$ has DS $\imp$ $H_{l+1}$ has DS }.
\vskip 0.15cm
Now we show that $-\Delta+V_{l+1}$ also has a DS using the 
	min-max principle. We start 
	with $-\Delta+V_{l+1}$ and get a bound in terms of $-\Delta+\sqrt{V_{l}}$.
We rewrite $V_{l+1}$ as

\begin{eqnarray}
 V_{l+1}=  \sum_{M_{1} \neq M_{2} \neq \ldots \neq M_{l+1}}(X^{a_{1}}_{M_{1}}\ldots 
 X^{a_{l+1}}_{M_{l+1}}f_{a_{1}\ldots a_{l+1}b_{l+2}\ldots 
 b_{L}}^{b})(X^{\hat{a}_{1}}_{M_{1}}\ldots 
 X^{\hat{a}_{l+1}}_{M_{l+1}}f_{\hat{a}_{1}\ldots 
 \hat{a}_{l+1}b_{l+2}\ldots b_{L}}^{b}) \cr 
    =\sum_{M} \left[\sum_{M_{1}\neq M_{2}\neq \ldots 
    \neq M}(X^{a_{1}}_{M_{1}}\ldots X^{a_{l}}_{M_{l}}f_{a_{1}\ldots 
    a_{l} c b_{l+2}\ldots b_{L}}^{b})(X^{\hat{a}_{1}}_{M_{1}}\ldots 
    X^{\hat{a}_{l}}_{M_{l}}
f^b_{\hat{a}_1\ldots\hat{a}_l\hat{c}b_{l+2}\ldots b_L})\right]X^{c}_{M} 
    X^{\hat{c}}_{M}  \nonumber 
\end{eqnarray}
and notice that it is a sum of harmonic oscillator potentials with a 
matrix coefficient not involving $X_{M}$. Let $\dis{\Delta_M=\sum_{c=0}^N}\frac{\partial^2}{(\partial X_M^c)^2}$, then for any $k$, $0<k<1$, we have

\begin{eqnarray}\label{II}
 -\Delta+V_{l+1}=(1-k)(-\Delta) + \sum_{M} [- 
 k\Delta_{M} + \qquad \qquad \qquad \qquad  \qquad \qquad \cr 
 \sum_{M_{1}\neq  \ldots \neq M_{l}\neq  M}(X^{a_{1}}_{M_{1}}\ldots  X^{a_{l}}_{M_{l}}f_{a_{1}\ldots a_{l}c\ldots b_{L}}^{b})(X^{\hat{a_{1}}}_{M_{1}}\ldots X^{\hat{a_{l}}}_{M_{l}}f_{{\hat{a_{1}}}\ldots {\hat{a_{l}}}{\hat{c}}\ldots b_{L}}^{b})X^{c}_{M}X^{\hat{c}}_{M}]  \cr 
     =(1-k)(-\Delta)+k\sum_M\left(-\Delta_{M}+\frac{1}{k}A^{(M)}_{c{\hat{c} }}X^{c}_{M}X^{\hat{c}}_{M}\right) \qquad \qquad \end{eqnarray}%\cr 
     \[\geq  (1-k)(-\Delta)+\sum_M \sqrt{k}\sqrt{\textrm{tr}\, A^{(M)}} 
     \geq  (1-k) \left[-\Delta + 
     \frac{\sqrt{k}}{(1-k)}\sqrt{V_{l}}\right] \]

By hypothesis $\sqrt{V_{l}}$ satisfies the sufficient condition of Theorem \ref{teo_mazya},  then so does $\frac{\sqrt{k}}{(1-k)}\sqrt{V_{l}}$, hence the 
righthand side of (\ref{II}) has a DS. It implies that 
$-\Delta+V_{l+1}$ also has a DS by Theorem \ref{min}. 
So the proposition is proved.
\end{proof}

\begin{rem}\emph{
If $-\Delta+\sqrt{V_l}$ has DS then $-\Delta+V_l$ has DS if $V_l$ is locally bounded i.e., for all compact set $K$ there exists $M_K>0$ such that $V_l(X)\le M_K$ for all $X\in K$, because of Lemma \ref{lema} and
\[\left(\int_{\Gd\setminus F}\sqrt{V_l}\,dX\right)^{2}\le \vol{\Gd\setminus F}\int_{\Gd\setminus F}\sqrt{V_l}\,dX\]}
\end{rem}
    With the latter proposition proved it is straightforward to 
    show that
    \begin{prop}
        $H_{l}$  has a discrete spectrum, for all $1\leq l \leq L$.
    \end{prop}
    \begin{proof}
        $H_{1}$ is the Hamiltonian for a harmonic oscillator, it then has a DS. One 
	makes use of the sequence proved in Proposition \ref{seq} to 
	conclude that $H_{l}$ has a DS.
    \end{proof}
    
 To show that the regularized $M5$-brane has a discrete spectrum 
 with eigenvalues tending to infinity, we need to prove first that the 
 $p$-brane has the maximum of its eigenvalues going to infinity, 
 since the proof of discreteness is not enough to prove a compact 
 resolvent.

\begin{prop}
        The spectrum of $H_{l}$ with $1\leq l \leq L$ has eigenvalues 
	$\lambda_{n}$ satisfying 
	\begin{equation}
	 \lambda_{n}\rightarrow\infty \qquad\textstyle{when}\qquad n\rightarrow \infty.
	    \label{eigen}
	\end{equation}
    \end{prop}
    \begin{proof}
        $H_{1}$ is the Hamiltonian for a harmonic oscillator , it  
	has a compact resolvent, hence its spectrum satisfies (\ref{eigen}). We now consider any of the potentials 
	$\sqrt{V_{l}}$ in the sequence of  Proposition \ref{seq}. Let us 
	denote it $V$. It is of the form
	\begin{equation}
	 V(X)= \sqrt{R^{n}}W(\hat{\theta }) \qquad 
	 \textrm{where}\qquad \|W(\hat{\theta })\|=1
	\end{equation}
	$R$, $\hat{\theta }$ are polar coordinates. We consider the 
	neighborhood $\Omega_{\epsilon}$ of zeros of $W(\hat{\theta })$, 
	i.e., 
	\begin{equation}
	 \Omega_{\epsilon}=\left\{\hat{\theta }:W(\hat{\theta }) < \epsilon 
	 \right\} \qquad\qquad\epsilon > 0.
	\end{equation}
	We then define $V_{\epsilon}(X)=\sqrt{R^{n}}W_{\epsilon}(\hat{\theta })$ where
	\begin{eqnarray}
	   W_{\epsilon}(\hat{\theta })  & = & W(\hat{\theta 
	   })\qquad\qquad X\in 
	   \textrm{complement of }\Omega_{\epsilon}   \\
	   W_{\epsilon}(\hat{\theta })  & = & \epsilon \qquad\qquad X\in \Omega_{\epsilon}.
	\end{eqnarray}

	For any $\epsilon>0$, $V_{\epsilon}(X)\rightarrow\infty$ as 
	$|X|\rightarrow\infty$, hence the spectrum of $H_{\epsilon}= 
	-\Delta+V_{\epsilon}$ satisfy (\ref{eigen}). 
	
	In fact, using the min-max theorem is easy to see  that if we define a Hamiltonian 
	$H_{well}= -\Delta+W$, where $W$ is a potential well $-c$ constant 
	in the region  inside a ball $S$ and $0$ outside it. And the 
	ball $S$ is taken in such a way that $V_{\epsilon}\geq c$ 
	outside it (this 
	is always possible since $V_{\epsilon}(X)\rightarrow\infty$ as 
	$|X|\rightarrow\infty$) then $V_{\epsilon}\geq c+W$ so that
	$\mu_{n}(H_{\epsilon})\geq c+ \mu_{n}(H_{well})$. Using the 
	fact that $\mu_{n}(H_{well})\geq -1$ with $n\ge N$, for some $N$, since 
	$W$ is a bounded potential of compact support, we get 
	$\mu_{n}(H_{\epsilon})\geq c-1$ if $n\geq N$ but since $c$ is 
	arbitrary $\mu_{n}(H_{\epsilon})\rightarrow\infty$ as 
	$n\rightarrow\infty$. So, the spectrum of $H_{\epsilon}$ 
	satisfy (\ref{eigen}).
	
	When $\epsilon \rightarrow 0$, the spectrum of  
	$H_{\epsilon}$ could become continuous but we already know 
	from Proposition \ref{seq} that it is discrete. It then 
	follows property (\ref{eigen}) also for the case when $\epsilon \rightarrow 0$.  
    \end{proof}
    \begin{prop}
     The Hamiltonian of the regularized bosonic $M5$-brane $H$ satisfies property \emph{(\ref{eigen})}.
    \end{prop}
    \begin{proof}
        We have $H > H_{4}$, from the above propositions $H_{4}$ 
	has a discrete spectrum satisfying property (\ref{eigen}). 
	Theorem  \ref{min} ensures then 
	$\mu_{n}(H)>\mu_{n}(H_{4})$ so the spectrum of $H$ also 
	satisfies (\ref{eigen}). $H$ has a compact resolvent in the 
	Hilbert space obtained by the completion of the subspace 
	generated by its eigenfunctions.
    \end{proof}
   
\section{Conclusions}

We showed  the discreteness
of the spectrum of the $M5$-brane and, in general, of $p$-brane theories.  The condition is
obtained from the Molchanov, Maz'ya and Shubin necessary and
sufficient condition on the potential of a Schr\"odinger operator
to have a discrete spectrum. The criteria is expressed not in
terms of the behaviour of the potentials at each point, but by a
mean value, on the configuration space.  The mean value in the
sense of Molchanov considers the integral of the potential on a
finite region of configuration space. It can be naturally
associated to a discretization of configuration space in the
quantum theory. We found that the mean value in the direction of
the valleys where the potential is zero, at large distances in the
configuration space, is the same as that of a harmonic oscillator. 
Also, using the min-max principle it was shown that the discrete 
spectra had eigenvalues running to infinity showing that their 
respective resolvent operators are compact.

\section*{Acknowledgments}

A. R. would like to thank Perimeter Institute for kind hospitality . 
Also, A. R. and I. M. are grateful to  E. Planchart and V. Strauss (USB), L. Boulton 
(Heriot-Watt U.), F. Cachazo (Perimeter Institute) and M. P. Garcia del Moral 
(Turin U.) for fruitful discussions. This work was supported by 
PROSUL under contract CNPq 490134/2006-8 and Decanato de 
Investigaciones y Desarrollo(DID-USB), Proyecto G-11.

\end{document}